# Open Research Data in Spanish University Repositories

Pablo Monteagudo-Haro*, Juan Jose Prieto-Gutierrez**

*Facultad de Ciencias de la Documentación, Universidad Complutense de Madrid https://orcid.org/0009-0003-2874-348X

**Facultad de Ciencias de la Documentación, Universidad Complutense de Madrid https://orcid.org/0000-0002-1730-8621

**ABSTRACT:** The current situation of open research data in Spanish university repositories is analyzed by means of twelve indicators that allow us to compare them with each other. The twelve self-developed indicators deal with research datasets and institutional policies linked to open access, as well as some of the key characteristics of the repositories. The methodology used consists of comparing the repositories of the different universities linked to REBIUN. The result has been that datasets in institutional repositories are scarce, and the situation is heterogeneous across the territory. This raises questions about future open access policies for research data in the country's main scientific institutions.

****************

1.  **INTRODUCTION**

In 2002 the foundations of open access were presented through the *Budapest Open Access Initiative* (Open Society Institute, 2002), which proposed the removal of barriers to scientific literature in order to accelerate research. It proposed institutional self-archiving and open access publishing as strategies for this. Subsequently, the *Bethesda Statement on Open Access Publishing* (Howard Hughes Medical Institute, 2003) and the *Berlin Declaration on Open Access to Knowledge in the Sciences and Humanities* (Max Planck Gesellschaft, 2003), both from 2003, sought to deepen open access by involving specific actors beyond researchers, such as libraries, publishers, institutions and funding agencies.

The consideration of open access implies the possibility of reading, downloading, copying, distributing, printing, searching or using for any legal purpose, without any financial, legal or technical barriers, apart from those that are inseparable from Internet access itself (Open Society Institute, 2002).

Directive (EU) 2019/1024 specifically talks about Open Access in the field of universities and research institutions. *Open Access* (open access), *Open Science* (open science) and *Open Data* (open data) are interrelated in the European university and scientific sphere.

Open Science refers to the dissemination of scientific knowledge in a free, accessible online and reusable way. Proposing an open science framework in research practices is understood under the promotion of open access to scientific research, reproducibility and open evaluation. In this openness context arises the need and interest in access to open research data, which are linked to the same scientific research, since research data allow the reproducibility and evaluation of the process, but also allow scientific reuse.

It is necessary to make a distinction between open data and open research data. The consideration of open data is related to the data produced by an institution in the development of its activity, which in the case of public administrations is linked to *Open Government*. Open government initiatives seek control over the actions of the government and the different institutions by any citizen or any interested social group. In this way, open government can be considered as the way to raise transparency with government information, openness in participation and citizen collaboration in decision-making and control mechanisms; it is a model that is presented within the standards of participatory democracy (Clabo and Ramos-Vielba, 2015). In this context, the possibility of reusing open data from public administrations is developed.

An example of this was carried out in the 2018 study on the data that the university as an institution generated around management data and academic data, following the policy standards for the reuse of public sector information (Martín González and Ríos Hilario, 2018). This study made explicit that the university produced three types of data: management, academic and research data (Martín González and Ríos Hilario, 2018: 117). This point of distinction is key to differentiate open data, which are usually linked to open government, and open research data, which are linked to scientific development.

Research data, whether open or not, are the objective evidence on the basis of which theses or research postulates are validated. Open research data are part of open science as long as an open dissemination policy is in place. In this scientific and research context, *Data Sharing* is considered as "the action of sharing with other colleagues the data files (or raw data) generated during the course of research" (Torres Salinas et al., 2010: 258), i.e., sharing the raw material generated during the course of the research. *Data Sharing* is linked to the open access philosophy because it favors openness and accessibility, in addition, it promotes a "second life" for those raw data, being able to reuse them for another purpose (Sixto-Costoya et al., 2019).

Sharing open research data allows to increase the impact and visibility of research, improve the reproducibility of science, the possibility of reuse of the same data as the original study in other subsequent studies, open the possibility of cost savings, encourage collaboration and increase credibility in the scientific system (Lyon, 2016).

In terms of scientific information management, the concept of *Data Curation* has generated a new area of responsibility for researchers, librarians and information professionals in the digital library environment (Heidorn, 2011), who are dedicated to the search, selection, characterization and continuous dissemination of the most relevant content from various information sources (Guallar and Leiva-Aguilera, 2013: 27), and, ultimately, of data (Tammaro, et al., 2019).

In the case of research in the European and Spanish context, the issue is mentioned around the transfer of results financed mostly with public funds (Law 37/2007; Law 14/2011; Law 18/2015 and Directive (EU) 2019/1024), referring to both publications and research data. The sense of this interest in sharing research results in an open way aims to improve the return on investment made by public institutions when funding research (Hernández-Pérez, 2016: 520) and to ensure open access to data for the public, governments and funding agencies (Stieglitz et al., 2020). In recent years, data deposition has been facilitated through several initiatives, such as the Zenodo repositories (created by OpenAIRE and CERN), Figshare, DataCite, etc.

In the context of the open access policies of Spanish universities, ten years ago three quarters had initiated some action in the framework of open science directed towards the creation of repositories or OpenCourseWare (OCW) courses (Abadal et al., 2013). OCW courses sought to make educational materials open and free to be consulted. Today, policy awareness has increased across the board towards data sharing (Gonzalez-Teruel et al., 2022). In addition, at present, the regulatory framework promotes the development of institutional or thematic open access repositories, own or shared linked to Spanish universities and research organizations to transfer research results to society (Law 17/2022 amending Law 14/2011), which mainly affects publications. Likewise, Article 12 of the Organic Law of the University System (2023) obliges teaching and research staff to make public the final version of their publications in scientific journals by depositing them in a repository. The Spanish legal mandate for research data is to be stored in a repository (Law 37/2007), without necessarily having to be in an institutional repository.

At the moment, the institutional repositories of Spanish universities house pre-prints and post-prints publications, among other documents, and following this same line of thought, it could be suggested that this same space could be used to house research data. Data that, in some way, are linked to those publications housed in institutional repositories, and that are developed in the academic institutions that maintain them. These data should comply with the international FAIR (Findable, Accessible, Interoperable and Reusable) principles, in order to make them easy to find, accessible, interoperable and reusable.

The link between research data and institutional repositories is raised under the interest that there could be between academic libraries and these data, since: "Research data repositories serve, among other purposes, to validate research results and, therefore, must be linked in some way to scientific publications where it is shown what those data were used for" (Hernández-Pérez and García-Moreno, 2013: 261).

There is no express legal mandate for the exclusive use of institutional repositories to house research data. The vast majority of Spanish universities use this platform to store their scientific production.

In order to specify this purpose, the limits of the research are developed in the following section on the objectives of the study.

## 2. OBJECTIVES

The object of the study is open research data in Spanish universities, specifically its situation in university repositories. The aim of the study is to analyze through a comparison the situation of institutional repositories in Spanish universities. To do this, we will look for relevant aspects that allow us to draw the current situation, and thus, to be able to point out some steps that might be necessary to achieve the objectives that the legal framework intends in relation to open data.

The following specific objectives are specified:

1. To analyze the descriptive characteristics of the institutional repositories of the Spanish universities that are part of the University Library Network REBIUN (Spanish Network of University Libraries).
2. To analyze previously established indicators to show the access to research data of Spanish universities in their repositories, and the digital context in which they are found.

## 3. METHOD

In order to respond to the stated objectives, a comparative method is used between the data extracted from the web observation of the institutional repositories of Spanish universities. The comparative method consists of analyzing the repositories of Spanish universities, establishing similarities and differences between the results obtained university by university, making it possible to compare the results, taking into account the ownership (public or private), the autonomous communities, or even the type of software of the repository, among many other aspects.

To obtain the results, a series of self-developed indicators (Table 1) were observed on the website of each of the institutions during the month of July 2022. These indicators are as follows:

Table 1. Indicators for the study of research data in institutional repositories

| | RESEARCH DATA IN INSTITUTIONAL REPOSITORIES | | |
|---|---|---|---|
| | INDICATORS | RESULTS | DESCRIPTION |
| i.01 | Hosting research data | Yes/No | Whether or not the institutional repository hosts research data. |
| i.02 | Location of data | Yes/No | Whether the repository itself has its own space for hosting research data. |
| i.03 | No. of data sets | Integer number | Indicates the number of datasets that are hosted in the repository. |
| | OPEN ACCESS POLICIES | | |
| | INDICATORS | RESULTS | DESCRIPTION |
| i.04 | General Open Access Policy | Yes/No | Whether the institution has a published open access policy document. |
| i.05 | Research Data Management | Yes/No | Whether the institution has a published document reflecting the research data management policy. |
| | CHARACTERISTICS OF THE REPOSITORIES | | |

|      | INDICATORS | RESULTS | DESCRIPTION |
|------|------------|---------|-------------|
| i.06 | Software | Program name | The type of software that supports the institutional repository is indicated. |
| i.07 | Metadata type | Metadata protocols | The different metadata protocols per institution are listed. |
| i.08 | Organize by metadata | Yes/No | It is indicated whether it is possible to organize the datasets according to the different metadata variables. |
| i.09 | Accessibility | Yes/No | Whether or not the documents in the datasets can be accessed. |
| i.10 | Licensing and copyright policies | License types | The type(s) of intellectual property license(s) that the repository uses on the datasets. |
| i.11 | Harvesters | Harvester names | Indicate to which collectors the repositories are linked. |
| i.12 | Publishing guide | Yes/No | Indicate whether the repository or library offers publication guidelines for research data. |

Source: own preparation

The main objective, current status of data in repositories, is proposed after developing the two specific objectives, analysis of descriptive characteristics and indicators, since prior collection of these results is necessary in order to be able to draw up a description, quantitative at first, of the data situation in 2022.

The first group of indicators, "Research data in institutional repositories", comprising indicators i.01, i.02 and i.03, and the second group of indicators, "Open access policies", indicators i.04 and i.05, apply to the entire sample of repositories of the universities linked to REBIUN, while the third group, "Repository characteristics", indicators i.06 to i.12, apply to the entire sample of repositories of the universities linked to REBIUN, while the third group, "Repository characteristics", indicators i.06 to i.12, apply to the entire sample of repositories of the universities linked to REBIUN. to i.12, apply to institutions that host data in their institutional repositories, or have a section for hosting data, since the analysis of repositories is of interest in relation to research data, and not to the evaluation of institutional repositories per se.

The analysis indicators differ from other proposals, such as REBIUN's work on repositories, Guía para la evaluación de Repositorios institucionales de Investigación (Barrueco et al., 2021). Its eight criteria are not replicated, since each of them extends in a very detailed analysis, and would exceed the objectives of this research. This REBIUN document is taken into account because it is the reference framework for Spanish academic institutions, but the indicators were also based on the text FAIREST: A Framework for Assessing Research Repositories (d'Aquin et al., 2023), since it specifically takes into account research data in repositories. It provides an analysis that is closer to the objective of the research, but, since it is from an international perspective, it has had to be adapted to the specific Spanish reality at some points.

The analysis sample focuses on the 76 Spanish universities that are members of CRUE and REBIUN, 50 public and 26 private. It should be noted that the University of Vigo and its Investigo repository have not been analyzed since it could not be accessed due to technical problems at the university during the information search period, making the final sample of 75 institutions.

## 4. RESULTS

The first aspect to be addressed is research data in institutional repositories.

### 4.1. RESEARCH DATA IN INSTITUTIONAL REPOSITORIES

The indicator "research data hosting" (i.01) is taken first because it will give a global view of the data in the 76 REBIUN universities. Data storage in all universities is indicated, and also, the data results are disaggregated in relation to whether the institution is public or private (Figure 1).

Figure 1: Percentage of REBIUN universities storing research data: total percentage, percentage of private institutions and percentage of public institutions.

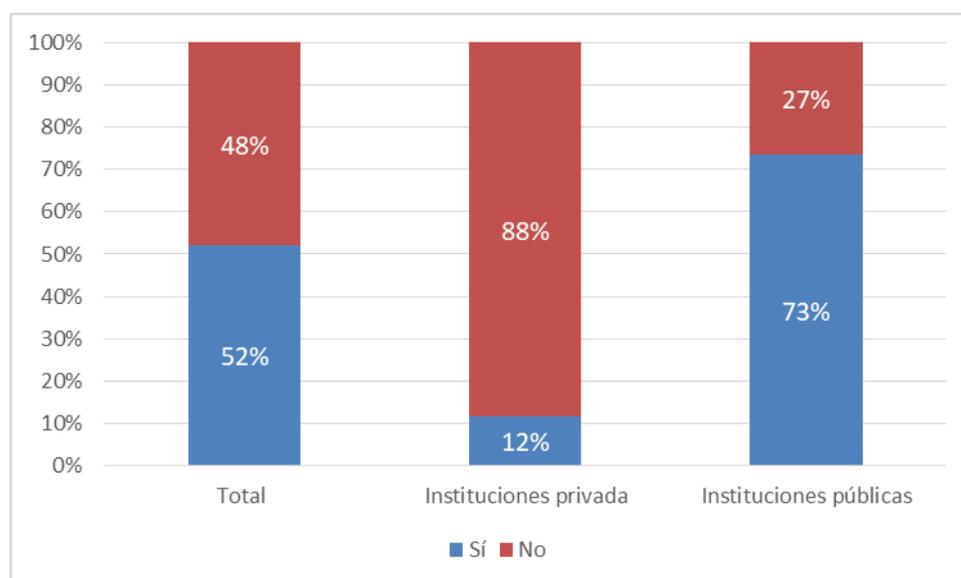

Source: own preparation

Public institutions lead in the storage of research data in repositories. If we look only at the group of public institutions, the result is 73%, which is 36 universities, while if we look at the total number of Spanish institutions the percentage drops to 52% (39 institutions). The number of public university repositories with research data is six times higher than that of private universities, which is 12% (3 institutions), these being the Universidad Camilo José Cela, Mondragon Unibertsitatea and Universitat Oberta de Catalunya.

By autonomous communities, Andalusia and Catalonia stand out, with eight institutions with research data each, and Madrid, with seven (Figure 2).

Figure 2: Number of REBIUN universities with research data in their repositories by autonomous community in Spain.

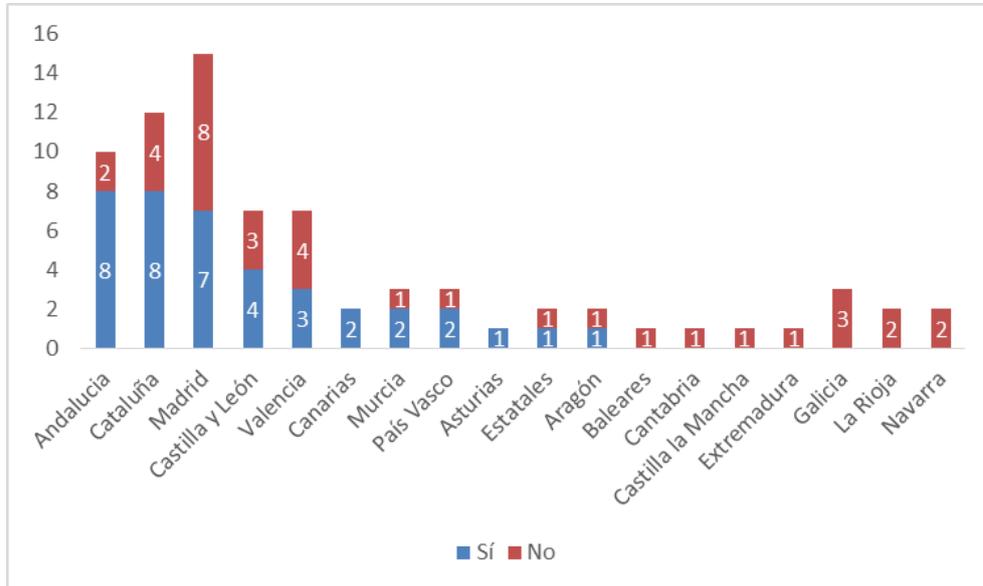

Source: own preparation

Figure 3 shows the five autonomous communities with the largest number of universities that have research data in their repositories, which are: Andalusia, Catalonia, Castile and Leon, Madrid and Valencia. The eight repositories in Andalusia represent 80% of the institutions in the region, while in Catalonia only 67%; in Castile and Leon it represents 57%, in Madrid, 47%, and in Valencia, 43%.

Figure 3: Percentage of universities in the community that have data in their repositories.

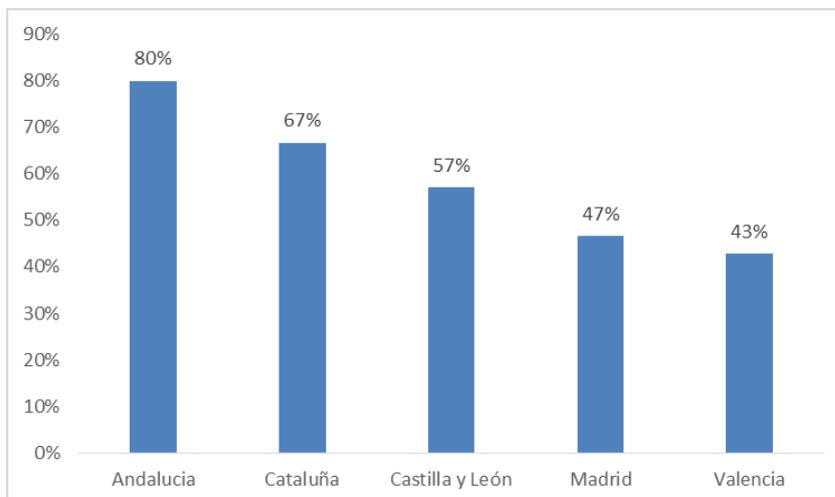

Source: own preparation

A comparison of Figure 2 and Figure 3 shows that: (a) Andalusia and Catalonia have the same number of universities committed to publishing data in their repositories (eight each), however, analyzing the percentage it seems that the Andalusian community is more committed, giving a result of 80% compared to 67%; (b) while Madrid is in fourth position if we look at the percentage, but in third position if we look at the total number of universities with data in their repositories.

The second indicator, "data location" (i.02), which defines whether the repository classifies research datasets in a specific section, makes it possible to identify the level of development of data accessibility in the repositories.

Figure 4 shows that the percentage of institutions that have a section for research data in their repositories is 56%, a higher percentage than the 52% of institutions that contain research data (Figure 1). This apparent difference is produced by three universities, Universidad Internacional de La Rioja, Universidad Francisco de Vitoria and Universidad Miguel Hernández de Elche, which have created specific sections for research data in their repositories, but do not contain any data as of July 2022.

Figure 4: Percentage of universities with a specific section for research data.

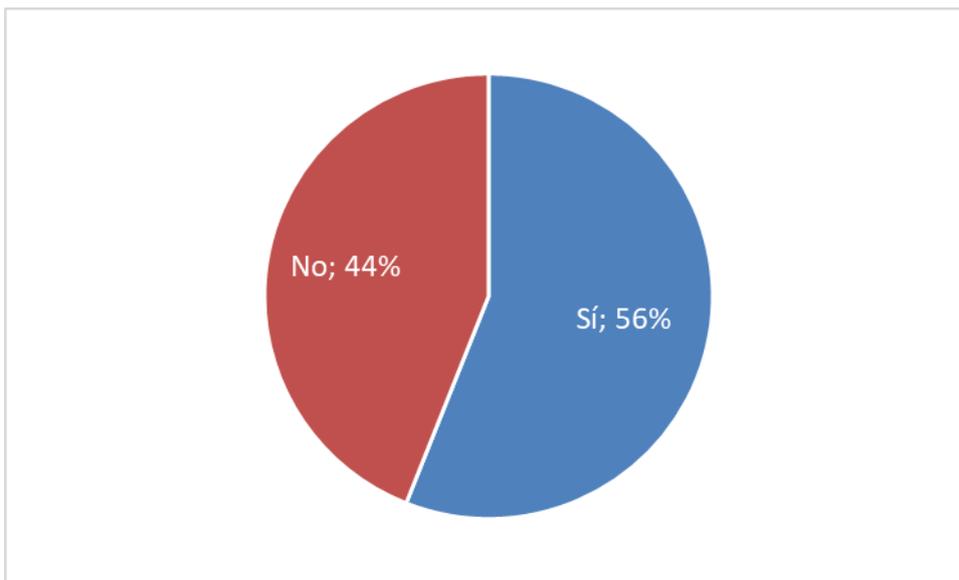

Source: own preparation

The indicator "number of datasets" (i.03) quantifies the number of datasets housed in each repository.

Some patterns that have emerged in reviewing the records should be noted:

(1) research data are published by recurring authors, the most representative case being that of the University of Zaragoza, where all datasets belong to the same person;

(2) in the sections created to store research data, in some cases other types of records are found, such as research support documents that are not strictly speaking research data.

Figure 5 shows that only four universities have stored more than 200 research data: the University of Alcalá (375), the Universitat Pompeu Fabra (268), the University Carlos III of Madrid (243), and the University of Zaragoza (222). It is noteworthy that among these four universities there are two from the community of Madrid.

Figure 5: No. of research datasets in the institutional repository by university

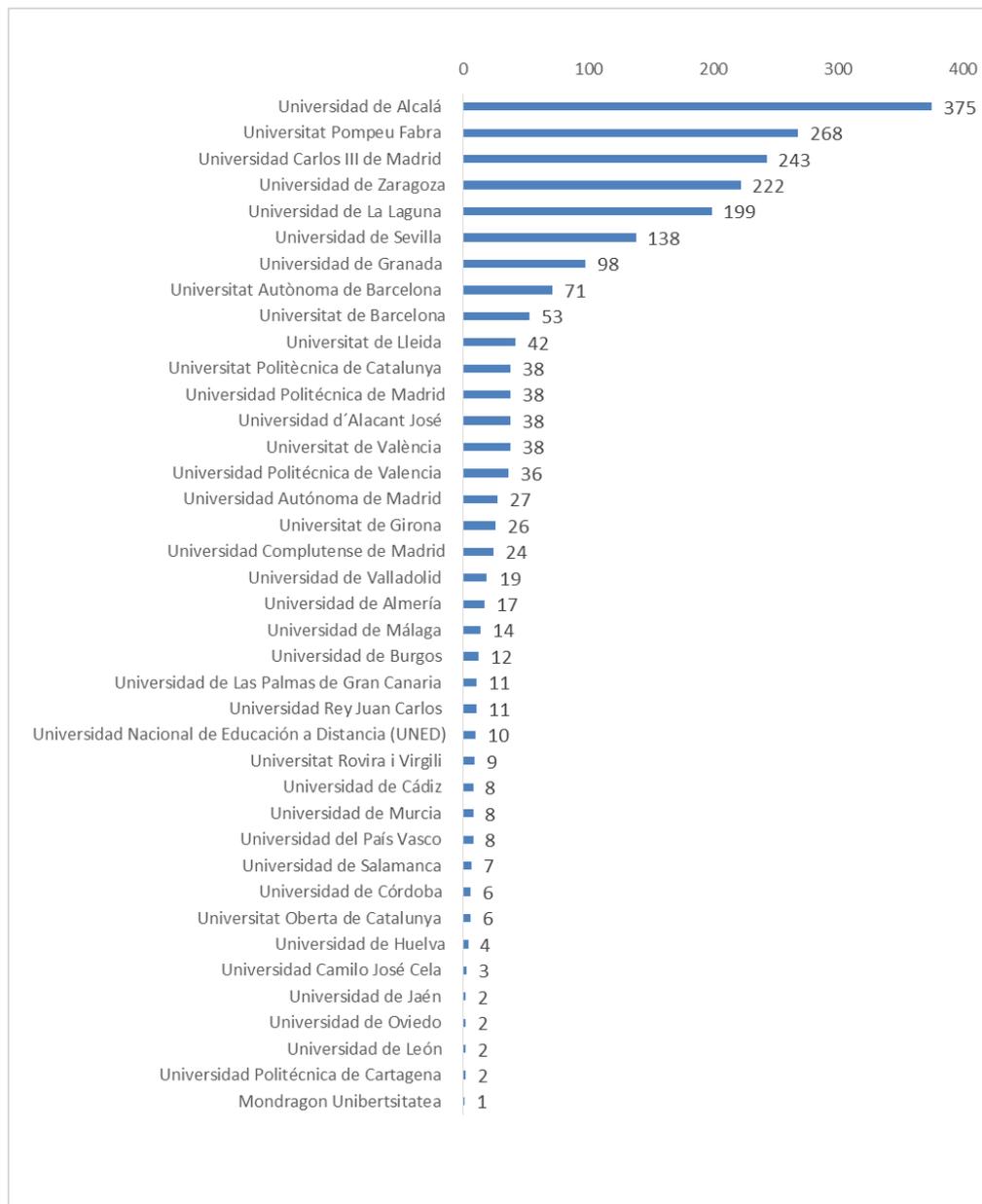

Source: own preparation

Figure 6 presents the cross-referenced data sets by autonomous community, where we can see that Madrid, Catalonia and Andalusia are at the top of the list.

Figure 6: Number of research data in institutional repositories by autonomous community.

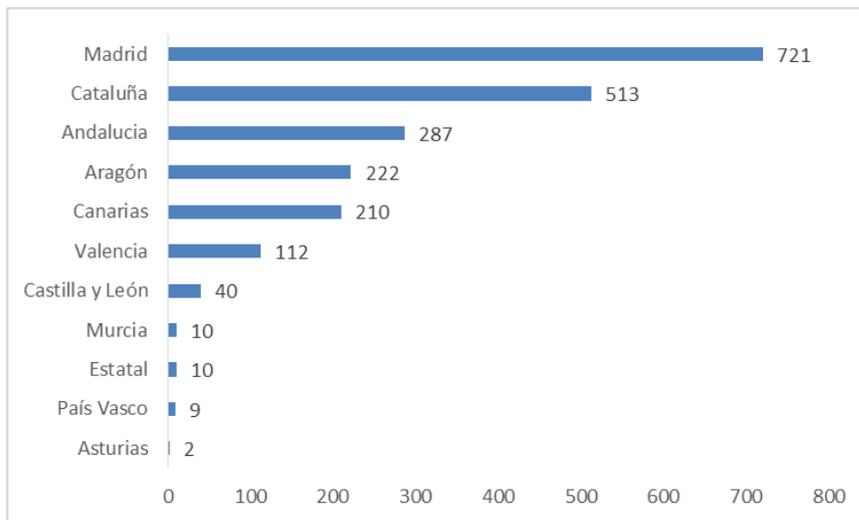

Source: own preparation

In Figure 2 we could see how Andalusia, Catalonia and Madrid had a very similar number of repositories containing research data: eight in Andalusia, eight in Catalonia, and seven in Madrid. On the other hand, Figure 3 shows the percentage of REBIUN universities with research data in their repositories by autonomous community: Andalusia led the Figure with 80%, Catalonia, with 67%, was second, and Madrid, with 47%, was in fourth position. If we contrast these data with the number of datasets by autonomous community, we can see that Madrid leads the number of data in repositories, followed by Catalonia and Andalusia, with the number of data in Madrid almost tripling the number of data in repositories in Andalusia. This shows that an autonomous community does not need to have a greater number of repositories to host research data in its repositories, but there is a correlation between the autonomous communities with the greatest number of repositories and the amount of data stored.

In Madrid, 36% of the total data are produced within the Madroño Consortium, made up of the Universidad Rey Juan Carlos, the Universidad Autónoma de Madrid, the Universidad Carlos III de Madrid, the Universidad Politécnica de Madrid, the Universidad de Alcalá and the UNED. These institutions represent 9% of the global REBIUN institutions.

**4.2. OPEN ACCESS POLICIES**

In this subsection, the link between the repository and open access is considered, for which two indicators are observed: whether the institution has a declared open access policy in general (i.04), and whether it has a specific policy for the management of research data (i.05).

Of the 75 institutions analyzed, 65% (49 institutions) have an open access policy (Figure 7). It is checked whether they host a public document indicating the institution's position on its accessibility and self-publication, i.e., a check has been made both in the repository and on the general web page of the university on whether it has this accessibility document, not only in the repository.

Figure 7: REBIUN Institutions that have an Open Access Policy

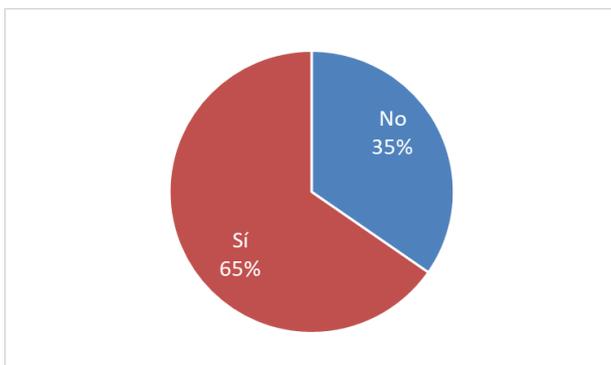

Source: own preparation

The indicator "research data management" (i.05) analyzes the institution's data management policy documents for research data, it allows to see the degree of development of the institutional protocol for the most appropriate treatment of research data for each institution.

Figure 8 shows that only 9% (seven institutions) of the universities analyzed have a specific policy for research data.

Figure 8: REBIUN Institutions with Research Data Management Policies

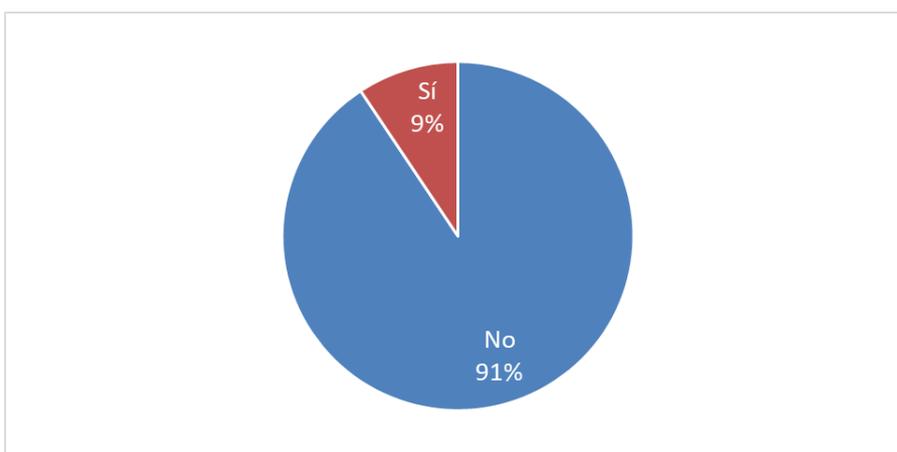

Source: own preparation

Figure 8 also shows that 91% (sixty-eight institutions) do not have a data management policy or a specific section on data management within their open access policies. Within this group, however, there is an internal division between organizations that do not cite research data in their open access policies, fifty-eight out of sixty-eight universities, and those that talk about research data as a part of scientific production along with articles and other scientific papers, ten of the sixty-eight institutions.

Six of the seven institutions that have access to this research data management document are part of the Madroño Consortium; the other university is the University of Malaga. The collaborative situation of the Madroño Consortium is not repeated in any other autonomous community. It could be considered a third way of sharing research data in addition to the two options mentioned above, which are: either to host the data in its repository or in an external repository.

## 4.3. CHARACTERISTICS OF THE REPOSITORIES

This section develops the indicators that allow us to analyze the current situation of the institutional repositories, giving rise to an analysis of the context of the university repositories where the housing of research data is observed.

Of the seventy-five repositories active in July 2022, forty-two have a section for research data, and thirty-nine contain some type of data. The group of repositories to be analyzed in this section are the forty-two repositories.

The indicator on software (i.06) identifies the program that supports the repository, with DSpace being the predominant software with 76.19% (32 repositories), followed by Dataverse, 14.29% (6 repositories), CDS Invenio is used by two repositories, and finally, Fedora and E-Prints are used in only one case each.

Regarding metadata (i.07), it is observed that all institutions with research data use Dublin Core (thirty-nine institutions), while six use METS, five institutions use MARC, and the same number of institutions use PREMIS, while only four use MODS, four EDM, and two MARCXML (Figure 9).

Figure 9. Number of repositories according to types of metadata used.

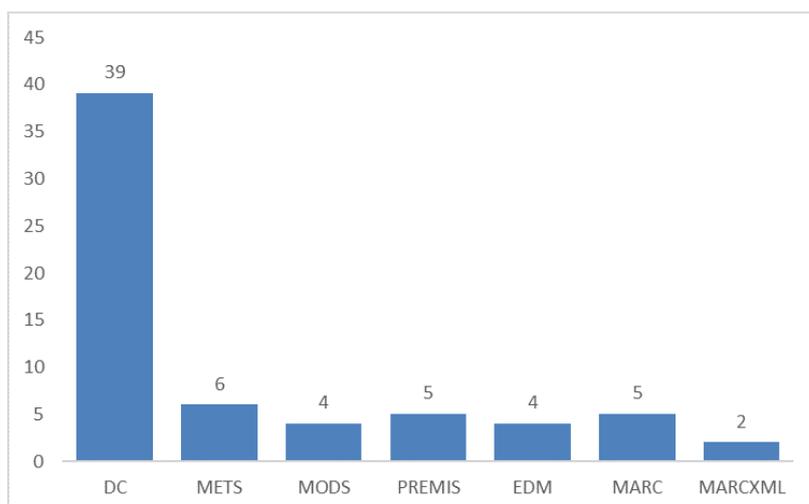

Source: own preparation

Seventy-six percent of institutional repositories with a section for research data only use a single metadata standard, which is Dublin Core.

Regarding organization by metadata (i.08), 97.6% of the 42 institutions allow searches to be organized according to the different metadata categories, with only one repository not allowing this.

The indicator on accessibility (i09), referring to whether or not the documents in the datasets can be accessed, of the thirty-nine repositories hosting data, thirty-eight allow access to all their research data directly without any restriction.

The status of copyright licenses associated with the *datasets* (i.10): thirty-seven repositories use some kind of Creative Commons license; and virtually all clearly indicate the license their data use.

Repository visibility is connected to the linkage of the repository to the different collectors (i.11): all repositories with research data are linked to the RECOLECTA collector, thirty-seven are linked to Google Scholar, thirty-four to BASE, thirty-two to OpenAIRE, nine to WorldCat and seven to Core. Most of the repositories are indexed in more than four collectors; this is the situation of twenty-eight out of thirty-nine repositories hosting data (72%).

The last indicator analyzes the existence of guidelines on open data publication (i.12), twenty-six of the repositories with data out of the thirty-nine repositories that host research data in their institutional repository have some guidelines of this type. It is noted that there are eight institutions that have no section for data or hosted research data, but have a data publication guide.

In general, looking at the characteristics of institutional repositories hosting data there is a strong homogeneity in most aspects: use of software linked to the Open Access movement, full accessibility to their data itself, ability to organize by metadata categories, something that is linked in many cases to the type of software, optimal ease of access to the type of license used in terms of copyright, as well as, there is a good result in linking to a minimum of four harvesters, which influences a good level of interoperability.

Overall, the characteristic results show a good level of functionality of institutional repositories hosting data, but two aspects are raised at this point to question this situation: the research datasets, which are low and heterogeneous, and the types of metadata. Since, in terms of metadata, 76.92% only present metadata in Dublin Core format.

## 5. DISCUSSION

The results analyzed show that the presence of research data in Spanish institutional repositories is limited; only four Spanish universities have more than two hundred research data stored in their repositories.

If we take into account that 65% of the repositories analyzed have open access policies, and that only 9% have developed specific policies for research data management, these results show a gap between the development of open access policies and policies for research data. It may be necessary to analyze other types of repositories to observe the behavior of research data in the Spanish context, whether they are thematic repositories or centralized repositories external to the institution.

However, this lack of specific policies reflects a lack of development of Data Sharing and Data Curation by universities and academic libraries.

One of the crucial aspects of Data Curation is the definition of policies and guidelines for research data management. Without established policies and practices, data is more likely to be inaccessible, lost, or inadequately managed. Back in 2016 IFLA initiated the Data Curator Project that sought to identify the key responsibilities of data curators and develop a glossary that should help to better define the profession and develop appropriate educational curricula. But the results show that most of the institutions analyzed lack specific research data management policies, highlighting the need for greater emphasis on Data Curation at the university level whatever the type of repository in which they are hosted.

There are two different situations observed in the autonomous communities with the most data, the Community of Madrid and Catalonia.

The situation of the Madroño Consortium shows that all its universities have research data management policies, being policies that are linked to the consortium. The data that are linked to the university institutions of this consortium are in a common repository managed by the Madroño Consortium, but that the universities themselves refer to it for the observation of their data, being the option that the institutional repositories allow. All this proposal would be linked to a clear practice of Data Curation by these universities linked to the consortium. However, this practice is not applied throughout the Community of Madrid, since not all universities are part of the consortium.

The situation of the Consorci de Serveis Universitaris de Catalunya (CSUC) with respect to research data management shows that there are Catalan universities that continue to host research data in their repositories, when the creation of a specific data repository has already been considered. The Catalan consortium had analyzed the different Catalan institutions to align policies (Alcalá and Anglada, 2019), which shows an interest towards a development of Data Curation. Two solutions were proposed from the observation of the situation: to use some existing repository, or to follow the policy adopted by some universities, which was to adapt their institutional repositories to deposit data. Finally, it was decided to create a specific repository for Catalan universities for this purpose, and it is a perfect example to illustrate the transition situation in which research Data Sharing in Spain finds itself.

The linkage between research, Data Sharing and Data Curation requires a more holistic approach to research data management in academic institutions, including the implementation of specific policies, the promotion of best practices specific to Data Curation and the adoption of more appropriate metadata standards. Since Data Curation is also closely related to the choice of metadata standards, it is significant that 76% of repositories with hosted research data only use Dublin Core.

## 6. CONCLUSIONS

Most institutional repositories are linked to open science, considering that they present open access policies. While most do not develop binding data management policies, two thirds present informative guidelines for data management. Taking into account that research data are linked to open science itself, and that scientific production is closely linked to universities, even if a deposit in other types of repositories is considered, Data Curation recommends universities and academic libraries to be the intermediary between the scientist and datasharing. Thus, there is room for improvement in these aspects of data management.

If we look at the institutional repositories that host data, in terms of their characteristics, there is a strong homogeneity in aspects that follow good practices recommended for this type of platform. The characteristic results show a good level of functionality of the institutional repositories that host data, but two aspects to improve are pointed out: the situation of the research datasets, and the types of metadata, most of them only use the Dublin Core format, also showing a possible path for improvement. Currently, there is open research data in some institutional repositories, but it is a very low number.

The datasets are presented in these institutional repositories that host data as quantitatively few in comparison with scientific publications, as well as, with research data that should be associated with Spanish research. In addition, a poor perspective of continuity is shown, if we attend to the data management policies. Six of the seven that have this policy are linked to a consortium, which is where these good practices are developed.

The situation in 2022 of open research data in repositories of Spanish universities suggests that the situation is one of transition, as if awaiting applied and practical policies that will establish in our context where and how research data will be stored, and specifically, what will happen to the research data that are currently in some institutional repositories.